\documentclass[aps,twocolumn,showpacs,amsmath,amssymb,pra,superscriptaddress,floatfix,longbibliography]{revtex4-1}
\usepackage{braket}
\usepackage{amsmath}
\usepackage{amssymb}
\usepackage{mathtools}
\usepackage{graphicx}
\usepackage{dcolumn}
\usepackage{bm}
\usepackage{multirow}
\usepackage{leftidx}
\usepackage{color}
\usepackage{float}

\usepackage{amsfonts}
\usepackage{eufrak}
\usepackage[german,english]{babel}

\begin{document}
\title{Simulations of sawtooth-wave adiabatic passage with losses}
\author{Johann Gan}
\affiliation{Department of Physics and Astronomy, Rice University, Houston,
Texas 77005, USA}
\author{M.E.~Pantalon}
\affiliation{Department of Physics, Kenyon College, Gambier, Ohio 43022, USA}
\author{F.~Robicheaux}
\email{robichf@purdue.edu}
\affiliation{Department of Physics and Astronomy, Purdue University,
West Lafayette, Indiana 47907, USA}
\affiliation{Purdue Quantum Science and Engineering Institute, Purdue
University, West Lafayette, Indiana 47907, USA}

\date{\today}

\begin{abstract}
The results of simulations of cooling based on
Sawtooth-Wave Adiabatic Passage (SWAP) are presented
including the possibility of population leaking to states outside of the
cycling transition. The amount of population leaking can be substantially
suppressed compared to Doppler cooling, which could be useful for systems
that are difficult to repump back to the cycling transition. The suppression
of the leaked population was more effective when simulating the slowing
of a beam than in cooling a thermal distribution. As expected,
calculations of the
leaked population versus branching ratio of spontaneous emission show
that the suppression is more effective for narrow linewidth transitions.
In this limit, using SWAP to slow a beam may be worth pursuing
even when the branching ratio out of the cycling transition is greater
than 10\%.
\end{abstract}

\maketitle
\section{Introduction}

The ability to laser cool atoms\cite{HJM99,CJF05} has enabled the exploration
of collective behavior in atomic gases, ultracold scattering, and
other effects. Methods used to laser cool atoms, as well as some variations,
have been used for molecules\cite{CDK09} and nano- and micro-scale
objects\cite{LNH12}. However, not all atoms, molecules, or nano-scale objects
can be effectively laser cooled with known techniques, which motivates
the search for new cooling methods.

Reference~\cite{NCB18} described a method for laser cooling called
Sawtooth-Wave Adiabatic Passage (SWAP) based on chirping
counter-propagating light waves. This method was proposed for cooling narrow
linewidth transitions. The basic idea is that Doppler shift of
the counter-propagating  light waves leads to a stimulated absorption
from the beam opposite the atom's velocity (due to the blue shift of
that beam) followed by a stimulated emission from the beam propagating
in the direction of the atom's velocity (due to the red shift of that
beam). If the linewidth is narrow, the spontaneous emission that 
occurs between
the stimulated absorption and emission will be small. This leads to
a momentum kick of $2\hbar k$, with $k$ the wavenumber of the light,
opposite the velocity of the atom. Experimental results on the Sr $^1$S$_0$
to $^3$P$_1$ transition with 7.5~kHz linewidth and
calculations from a simple theoretical model
supported both the possibility of cooling and the interpretation
of the mechanism.

Reference~\cite{BNC18} described in more detail the
simple theoretical model for this process
and presented the results from several simulations. Reference~\cite{PMB19}
performed SWAP cooling of Dy using a transition at 626~nm with a
136~kHz linewidth. This transition is $\sim 20\times$ broader than that
in the demonstration on Sr. Reference~\cite{SPH19} described experimental
results for SWAP cooling of Sr in a magneto-optical trap. 
These studies\cite{NCB18,BNC18,PMB19,SPH19,LCB19} showed that SWAP
gives a more rapid reduction of the kinetic energy compared to Doppler
cooling, although the final temperature that can be reached is several
times hotter than Doppler cooling.
Finally, as first suggested
in Ref.~\cite{NCB18}, SWAP cooling is promising for molecules
because the loss of population to states outside of the cycling transition
can be reduced due to the relative suppression of spontaneous emission.
Reference~\cite{LCB19}
calculated SWAP cooling for diatomic molecules in a magnetic field
and did find a reduction in the population into leakage channels.

In this paper, we perform calculations for a more simplified case than
Ref.~\cite{LCB19} to understand the role of leakage
during SWAP cooling. For the model below,
there is only one leak state and it can not
be connected to the excited state by a laser transition. While not
quite as realistic as Ref.~\cite{LCB19}, 
this model allows for a more transparent interpretation
of the results of the cooling when population can leak from the
cycling transition. In particular, we present results on the performance
of SWAP cooling as a function of branching ratio, ${\cal B}$, into
the leak state for various spontaneous decay
rates.

SWAP cooling for small ${\cal B}$ is important for molecules,
but larger branching ratios might also be interesting.
Part of our motivation was to determine whether SWAP
is worth pursuing when this branching ratio is greater than 1/2. For example,
laser cooling of antihydrogen, $\bar{\rm H}$, on the 1S-2P transition
was predicted\cite{DFR13}
to give an average final energy $\bar{E}_f\simeq 30$~mK compared to
a Doppler temperature of $\sim 2$~mK. Because of the long lifetime
of the 2S state, one could attempt SWAP cooling on the 2S-3P or
2S-4P transition to obtain colder
$\bar{\rm H}$ which would improve, for example, the measurement of
the 1S-2S transition\cite{ALP18}.
Unfortunately, simulations below suggest that
the short lifetime of these states
and the unfavorable branching ratio preclude cooling $\bar{\rm H}$ by SWAP.

The results presented below suggest that SWAP with leaks to other
states is more effective for
slowing a beam of particles than for cooling. As expected, the
quality of the SWAP cooling
increases as the spontaneous decay rate decreases. However, for very narrow
lines, SWAP might be useful
even for branching ratios, ${\cal B} > 0.1$.

\section{Theoretical model}

All of the calculations used the model introduced in Ref.~\cite{BNC18}
with a couple modifications which will be
explicitly noted.

The system is an atom with center of mass motion constrained
to one dimension and with 3 internal
states instead of the 2 internal states of
Ref.~\cite{BNC18}.
The translational motion is represented by momentum
eigenstates which are stepped in units of the photon momentum,
$\hbar k$ with $k$ the photon wavenumber.
The two internal states treated in Ref.~\cite{BNC18} are
a ground state $|g\rangle$ and an excited state $|e\rangle$.
To incorporate the possibility of population leaking to other states, the
calculations below include a third state $|l\rangle$. For simplicity,
the counterpropagating lasers cause transitions between the ground
and excited state but do not connect the excited and leak
state. The state of the atom
is specified by its internal state and its momentum state which is given
in multiples of the photon momentum. For example,
$|e,-6\hbar k\rangle$ represents the atom being in the
excited state {\it and} in the $-6$ momentum state. For conciseness,
we will use the symbol $n_p$ to refer to the momentum state; in the
example, $n_p=-6$.

The master equation
\begin{equation}
\frac{d\hat{\rho}(t)}{dt}=\frac{1}{i\hbar}[\hat{H}(t),\hat{\rho }(t)]+
{\cal \hat{L}}(\hat{\rho})
\end{equation}
determines the evolution of the atom through the time dependence
of the density matrix, $\hat{\rho}$. The time dependent
Hamiltonian, $\hat{H}(t)$, leads to coherent evolution of the system
from the stimulated absorption or emission of a photon and
the associated recoil.
The Lindblad superoperator, ${\cal \hat{L}}(\hat{\rho})$,
models the incoherent evolution from the spontaneous emission and
{\it its} associated recoil.

Within the rotating wave approximation, the time dependent Hamiltonian is
\begin{equation}
\hat{H}(t)=\frac{\hat{p}^2}{2M}-\frac{\hbar}{2}\Delta (t)\hat{\sigma}_z
+\frac{\hbar}{2}\Omega_s W(t)\cos (k\hat{z})\hat{\sigma}_x
\end{equation}
where $\Delta (t)$ is the time dependent detuning,
$\hat{\sigma}_z =|e\rangle\langle e|-|g\rangle\langle g|$,
$\Omega_s$ is the standing wave Rabi frequency, the $W(t)$
is a window function not used in Ref.~\cite{BNC18} but defined below,
$\hat{\sigma}_x=|e\rangle\langle g| + |g\rangle\langle e|$,
and
\begin{equation}
\cos (k\hat{z})|n_p\hbar k\rangle = 
\frac{1}{2}|(n_p-1)\hbar k\rangle +\frac{1}{2}|(n_p+1)\hbar k\rangle .
\end{equation}
The time dependent detuning has a sawtooth profile that goes from
$-\Delta_s/2$ to $\Delta_s/2$ with linear ramp, $\alpha$:
$\Delta (t) = -\Delta_s/2 +\alpha t$ with $t=0$ defining the start
of the ramp. The duration of the ramp, $T_s=\Delta_s/\alpha$, will
be used below.

For $\hat{H}(t)$,
the only difference from Ref.~\cite{BNC18} is that we use a windowing
function to turn the standing wave on and off. The windowing function
reduces the ringing
that results from instantaneous changes in the detuning by smoothly
going to 0 at the beginning and end of the ramp. In all of the calculations,
we used $W(t)=\exp [-36 (t-T_s/2)^8/(T_s/2)^8]$ where $0\leq t\leq T_s$,
but almost any smooth
function which is flat during the middle part of the ramp will lead
to similar results.

The Lindblad superoperator is somewhat more complicated than in
Ref.~\cite{BNC18} due to the branching ratio to the leak state. We
will use $\gamma$ for the total decay rate of the excited state and
${\cal B}$ as the branching ratio to the leak state, $|l\rangle$, which is
part of the cycling transition; $1-{\cal B}$ is the branching ratio to
the ground state, $|g\rangle$. The Lindblad superoperator is given by
\begin{eqnarray}
{\cal L}(\hat{\rho})&=&-\frac{\gamma}{2}(|e\rangle\langle e|\hat{\rho}
+\hat{\rho}|e\rangle\langle e| -2[\frac{3}{5}\hat{\rho}_p\nonumber\\
&\null&+\frac{1}{5}e^{ikz}\hat{\rho}_pe^{-ikz}
+\frac{1}{5}e^{-ikz}\hat{\rho}_pe^{ikz}]
\end{eqnarray}
where
\begin{equation}
\hat{\rho}_p=(1-{\cal B})\hat{\sigma}_g^-\hat{\rho}\hat{\sigma}_g^+
+ {\cal B}\hat{\sigma}_l^-\hat{\rho}\hat{\sigma}_l^+
\end{equation}
with $\hat{\sigma}_g^-=|g\rangle\langle e|=\hat{\sigma}_g^{+\dagger}$ and
$\hat{\sigma}_l^-=|l\rangle\langle e|=\hat{\sigma}_l^{+\dagger}$.
The Lindblad superoperator of Ref.~\cite{BNC18} results when
${\cal B}=0$.

\section{Basic parameters}

As noted in Ref.~\cite{BNC18}, the dynamics of this master equation
can be scaled, which means the outcomes are determined by scaled parameters.
We will define the parameters in terms of the total decay rate of the
excited state, with the scaled parameters being denoted by an over-tilde.
The important parameters are: $\tilde{\Omega}_s=\Omega_s/\gamma$,
$\tilde{\alpha}=\alpha /\gamma^2$, $\tilde{\Delta}_s=\Delta_s/\gamma$,
and $\tilde{\omega}_r=\omega_r/\gamma$ where $\hbar\omega_r = (\hbar k)^2/(2M)$
is the recoil energy.

As discussed in Ref.~\cite{BNC18}, there are dimensionless ratios that
are important for the effectiveness of SWAP. The adiabaticity parameter
$\kappa = \tilde{\Omega}_s^2/\tilde{\alpha}$ determines whether the
 the $g\leftrightarrow e$ transition is adiabatic
or diabatic. At the Landau-Zener level, the probability for an adiabatic
transition is $P=1-\exp (-\pi\kappa /2)$. The range of the sweep has to
be large enough to contain the Doppler shifted resonances plus a bit
extra to accommodate the transients at the beginning and end of the
sweep: $\Delta_s> 4 k |v|$. The amount of time spent in the excited
state should be much less than the lifetime of the excited state:
$2 k |v|/\alpha \ll (1/\gamma)$. Finally, the resonances should be
separated: $\Omega_s/2 < |k v - 2\omega_r|$.

\section{Results and Discussion}

The results from three different cases are presented in
this section. The case where the population is confined to a cycling
transition is briefly treated before the more complicated 3 state
system. For the purpose
of obtaining an absolute energy in some plots, we fixed the wavelength to
be $\lambda = 689.5$~nm and kept $\hbar\omega_r $
fixed at $0.36$~$\mu$K times Boltzmann's constant, $k_B$.

References~\cite{NCB18,BNC18,PMB19,SPH19,LCB19}
suggested that SWAP would be useful for the case where there is a
leak in the cycling transition to other states. The idea is that
the stimulated emission step would suppress the spontaneous emission
into the leak state(s), $|l\rangle$. The results below give an
indication of how effective this might be.

\subsection{No leak, ${\cal B}=0$: steady state vs $\tilde{\omega}_r$}

\begin{figure}
\resizebox{80mm}{!}{\includegraphics{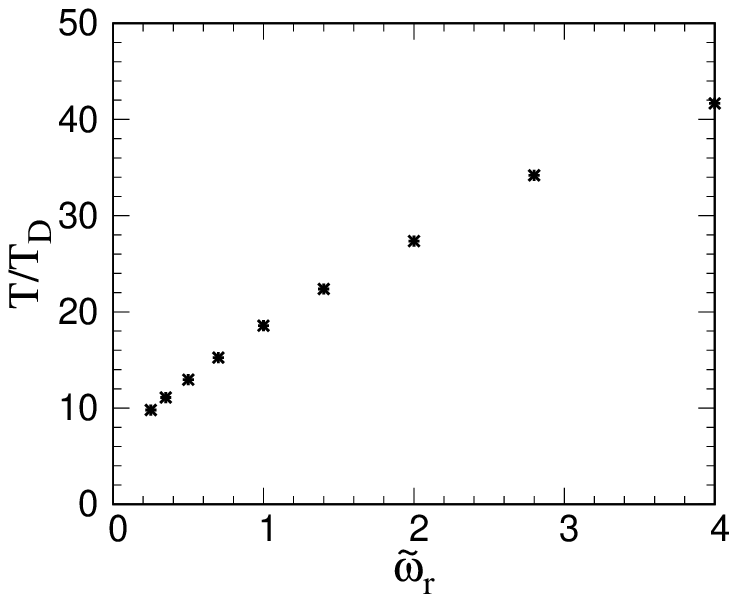}
\includegraphics{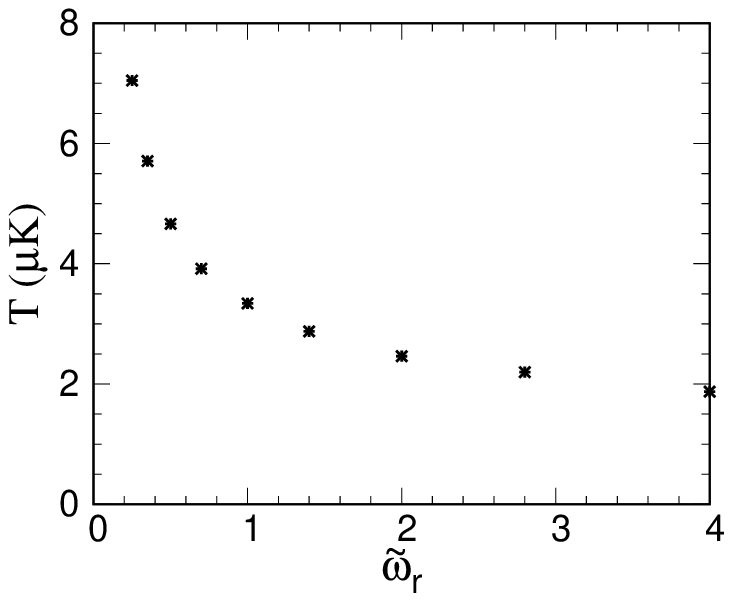}}
\caption{\label{FigNoLeak}
The asymptotic temperature, after many sweeps, as a function of the
scaled recoil frequency, $\tilde{\omega}_r$. The temperature is divided
by the Doppler temperature, $k_BT_D=\hbar\gamma /2$, in the left
plot. For this
case, there is no loss to state $|l\rangle$. In both cases,
we have kept the photon wavelength fixed at $\lambda = 689.5$~nm.
For both plots the temperature is defined as
$T\equiv 2\langle KE\rangle /k_B$.
}
\end{figure}

Before treating cases where population can leak into
state $|l\rangle $, it is worthwhile to examine how the excited state
linewidth affects the effectiveness of the SWAP procedure. We did
this by investigating
the steady state temperature reached using SWAP, as a function of the
scaled recoil frequency, $\tilde{\omega}_r=\omega_r/\gamma $.
SWAP was proposed as a method
for cooling atoms with narrow linewidths. This corresponds to larger
scaled recoil energies.

For these calculations, we fixed the range $\tilde{\Delta}_s=240$ and
set the ramp rate at $\alpha = \Omega_s^2/2$ (i.e.~$\kappa = 2$) which
gives a Landau-Zener probability of $P\simeq 0.96$. The initial momentum
distribution was started as a thermal distribution at low temperature
and the SWAP procedure was iterated several
times until the average kinetic energy
reached a limiting value. After each SWAP, we allowed the decay of any
population in the excited state as described in Ref.~\cite{BNC18}.
We also set the coherence between different momentum states to be
zero after each SWAP.
For each $\tilde{\omega}_r$, the Rabi frequency, $\tilde{\Omega}_s$,
was varied in steps of 1
to approximately find the lowest steady state energy. Different
values of $\tilde{\omega}_r=\varepsilon_r/(\hbar\gamma )$ were obtained
by varying the decay rate, $\gamma$.

The data in Fig.~\ref{FigNoLeak} shows the scaled average kinetic energy in
steady state versus the scaled recoil frequency. As was found in
Refs.~\cite{NCB18,BNC18,PMB19,SPH19,LCB19},
SWAP cooling does not achieve a temperature as low as that from
Doppler cooling; for the range plotted in Fig.~\ref{FigNoLeak}, the
asymptotic temperature was more than $\sim 10X$ the Doppler temperature
for every $\tilde{\omega}_r$.
However, as expected from Refs.~\cite{NCB18,BNC18,PMB19,SPH19,LCB19},
the lowest temperature is
achieved when the decay rate, $\gamma$, is smallest (i.e. when
$\tilde{\omega}_r$ is largest). Finally, as was seen
in Refs.~\cite{NCB18,BNC18,PMB19,SPH19,LCB19}, the simulations
showed that the kinetic energy is extracted from the atom much
faster using SWAP than using Doppler cooling.

\subsection{One sweep with loss}

The case where population can leak to state $|l\rangle $ is somewhat
more complicated because the loss grows with each SWAP and the
population that is lost remains at the same energy. To understand the trends
for SWAP with a leak out of the cycling transition, it is sufficient
to understand the results from one SWAP.

\subsubsection{Single initial velocity}\label{SecVel}

\begin{figure}
\resizebox{80mm}{!}{\includegraphics{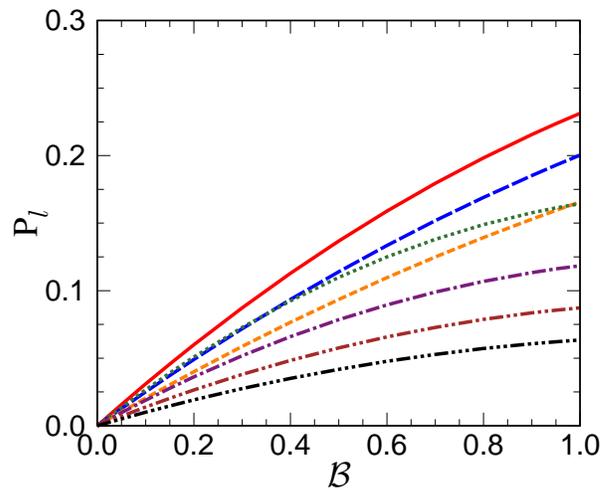}}
\caption{\label{FigLeakV}
The probability, $P_l$, for the atom to leak to the state $|l\rangle $
after one SWAP as a function of the branching ratio, ${\cal B}$,
to the leak state.
The solid (red) line
is for $\tilde{\omega}_r=0.25$, the medium dash (blue) line is for 0.35,
the short dash (orange) line is for 0.5, the 
the dotted (green) line is for 0.7, the dash-dot (purple) line is
for 1.0, the dash-dot-dot (maroon) line is for 1.4, and the
dash-dot-dot-dot (black) line is for 2.0. For all calculations, the atom
starts in the ground state with $p=20\hbar k$.
}
\end{figure}

In this section, the initial state has all of the population in one
particular momentum state. One sweep is performed. This situation
mimics that of slowing a beam of atoms. The quantities
of interest are the amount of population lost into state $|l\rangle $
and the change in energy of the states. In all of the calculations,
the atom starts in the ground state with $p=20\hbar k$.

There are several
parameters that control the population lost into the leak state and the
energy removed during the SWAP.
We chose to study how the population in the leak state and the final energy
varied with the branching ratio, ${\cal B}$, to spontaneously decay
to the leak
state, $|l\rangle$, for a few scaled
recoil frequencies, $\tilde{\omega}_r$. The
$\tilde{\omega}_r$ was stepped by the factor $\sim 1.4$ to show a range
of behavior. The range of
the sweep, $\Delta_s$, was scaled for each $\tilde{\omega_r}$ so
that $\tilde{\Delta}_s = 240\tilde{\omega}_r$. For each $\tilde{\omega}_r$,
the $\tilde{\Omega}_s$ was varied to give the best slowing for ${\cal B}=0$
and was then fixed for all other values of the branching ratio.
The ramp rate was chosen so that $\alpha = \Omega_s^2/2$ (i.e. $\kappa =2$)
which gives a Landau-Zener probability of $P\simeq 0.96$. A 
slower ramp rate would give a larger Landau-Zener probability but at
the cost of spontaneous emission between the stimulated absorption and
stimulated emission steps. After
the SWAP, the excited state population was allowed to decay as in
Ref.~\cite{BNC18}.

The parameters in the simulation gave a final energy after
one SWAP that ranged from
$0.82\times$ the initial energy for $\tilde{\omega}_r=2$ to $0.84\times$
the initial energy for $\tilde{\omega}_r=0.25$. For a perfect SWAP, the
momentum should go from $20\hbar k$ to $18\hbar k$ meaning the
expected final energy is $(18/20)^2=0.81\times$ that of the initial
energy. This indicates that the SWAP procedure is effective at slowing
the atoms for $n_p\sim 20$,
roughly independent of the decay
rate. As an example, for $\tilde{\omega}_r=1$,
89\% of the population finishes in the $18\hbar k$ momentum state
when ${\cal B}=0$.

Figure~\ref{FigLeakV} shows the population that leaks to the 
state $|l\rangle$
as a function of ${\cal B}$. As expected, there is no population leak for
${\cal B}=0$, and the population in $|l\rangle$ increases
with the branching
ratio. For small branching ratio into the leak state, ${\cal B}$,
the leaked population, $P_l$, is proportional to ${\cal B}$,
but increases slower than linear for larger ${\cal B}$.
The slope of $P_l$ for ${\cal B}\sim 0$ determines the effectiveness
of the SWAP procedure at small
branching ratios: smaller slope means less loss and a greater
effectiveness. Although the ${\cal B}\sim 0$ is important, even
the extreme case ${\cal B}\sim 1$ (i.e.~the branching ratio of spontaneous
emission into the leak state is $\sim$100\%), has less than 1/4 of the
population leaking into state $|l\rangle$
during a SWAP for the parameters in Fig.~\ref{FigLeakV}. As an example,
for $\tilde{\omega}_r=1$, the ${\cal B}\sim 1$ population loss is 12\%
per sweep. If this loss were the same for successive $n_p$, there would
be approximately 70\% population loss after 10 SWAPs. Similarly,
the $\tilde{\omega}_r=1$, ${\cal B}= 1/2$ case gives 8\% loss per
SWAP which is a factor of $\simeq 6$ suppression of the loss that would
occur for Doppler cooling.

With one exception, there is a clear trend of decreasing population leak
as $\tilde{\omega}_r$ increases. This is understandable because an
increased scaled recoil energy means a smaller decay rate, which should
make the SWAP procedure more effective: there is more stimulated emission
back to $|g\rangle$ and less spontaneous emission which could lead to
either $|g\rangle$ or $|l\rangle$. As an example, the
$\tilde{\omega}_r = 1$ case (dash-dot (purple) line)
used $\tilde{\Omega}_s=33$ and had $P_l\simeq {\cal B}/5$ for
${\cal B}\simeq 0$. For this case, the loss is $\simeq 5\times$
smaller than the best case using Doppler cooling. Even the worst case
shown ($\tilde{\omega}_r=0.25$ solid (red) line) has a loss for
${\cal B}\simeq 0$ that is $\simeq 3\times$ smaller than the best
case using Doppler cooling. The exception to the trend is
the $\tilde{\omega}_r=0.7$ (dotted (green) line). This case also had
an anomalous value for $\tilde{\Omega}_s$ compared to the trends observed
for the other $\tilde{\omega}_r$. We do not know why this case is different
from what we expected.

\subsubsection{Thermal distribution}

\begin{figure}
\resizebox{80mm}{!}{\includegraphics{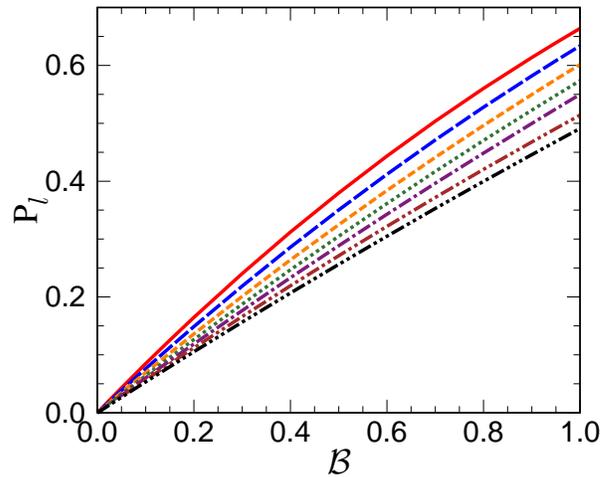}}
\caption{\label{FigLeakT}
The probability, $P_l$, for the atom to leak to the state $|l\rangle $
after one SWAP as a function of the branching ratio, ${\cal B}$,
to decay to the leak state, $|l\rangle$.
The line types are the same as Fig.~\ref{FigLeakV}.
For all calculations, the atom
starts in a thermal distribution of momenta as described in the
text.
}
\end{figure}

In this section,
we present the results for the effect of one SWAP when the initial
distribution of momenta is thermal. The initial distribution was
chosen to be proportional to $\exp (-[n_p/14]^2)$ which corresponds
to a temperature, $k_BT= 98\hbar\omega_r$. This temperature was chosen
to give similar SWAP parameters to Sec.~\ref{SecVel}. As with the previous
section, we performed calculations for several $\tilde{\omega}_r$,
the range of the sweep was $\tilde{\Delta}_s = 240\tilde{\omega}_r$,
and we varied the $\tilde{\Omega}_s$ for each $\tilde{\omega}_r$ to obtain
the best cooling for ${\cal B}=0$.

There was not a substantial difference
in the amount of energy removed in one SWAP for different
$\tilde{\omega}_r$ at ${\cal B}=0$. This trend is similar to
that in the previous section. The
final average energy somewhat increased with decreasing $\tilde{\omega}_r$:
largest for $\tilde{\omega}_r=1/4$ (approximately
$0.83\times$ the initial average energy) and smallest for 
$\tilde{\omega}_r=2$ (approximately
$0.78\times$ the initial average energy).

The population that leaked to state $|l\rangle$ after one SWAP is
plotted in Fig.~\ref{FigLeakT} for several different $\tilde{\omega}_r$.
The line types are the same as for Fig.~\ref{FigLeakV}. The amount
of population that leaks out of the cycling transition is much larger
than for the previous section where only one momentum component is
initially occupied. This is because the thermal distribution has
several momentum components, and the SWAP parameters were chosen to give
an overall efficiency. However, parameters that work well for large momenta
are not so good for small momenta and vice versa. The best slope
for small ${\cal B}$ is for $\tilde{\omega}_r=2$ and is $\sim 2\times$
smaller than would result from spontaneous emission from the excited
state. Unlike the previous section, the interpretation of this slope
is ambiguous because it is not clear how many photons were absorbed.

\begin{figure}
\resizebox{80mm}{!}{\includegraphics{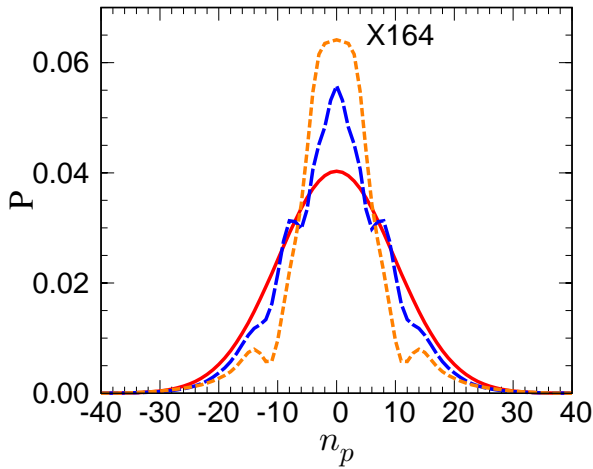}}
\caption{\label{FigPDis}
The probability, $P$, for the atom to be in the momentum state
$n_p\hbar k$. The solid (red) line is the initial momentum
distribution. The medium dash (blue) line is the momentum distribution
when the atom finishes in the $|g\rangle$ state for $\tilde{\omega}_r=1$
and the short dash (orange) line is when the atom finishes
in the $|l\rangle$ state. This line has been scaled to give the
same area as the medium dash (blue) line.
The branching ratio to spontaneously decay back to $|g\rangle$
is $1-{\cal B}=0.99$ for these results.
}
\end{figure}

The momentum distribution after SWAP, Fig.~\ref{FigPDis},
indicates that the atoms that leak into the state $|l\rangle$ are those
with smaller energy.
In this calculation, $\tilde{\omega}_r=1$ and the branching ratio
back to $|g\rangle$ is $1-{\cal B}=0.99$. For this case, 0.605\%
of the atoms leak to the state $|l\rangle$.
The solid (red) line shows the initial thermal distribution of
the atoms. The medium dash (blue) line shows the momentum distribution
after the SWAP for the atoms that finish in the $|g\rangle$ state.
The increase in the distribution for small $|n_p|$ shows that there
has been cooling during the SWAP,
even with the loss. The short dash (orange) line
is the momentum distribution for atoms that leaked to $|l\rangle$.
This distribution has been scaled so the integral would be the same
as for the $|g\rangle$ state. This distribution is even more strongly
peaked at small $|n_p|$ which indicates that the atoms that leak
preferentially have smaller energy. This trend is plausible because
choosing $\tilde{\Omega}_s$ to give the largest energy decrease indicates
the SWAP is most efficient for the states with larger $|n_p|$.

\subsection{Relative efficiency with loss}

\begin{figure}
\resizebox{80mm}{!}{\includegraphics{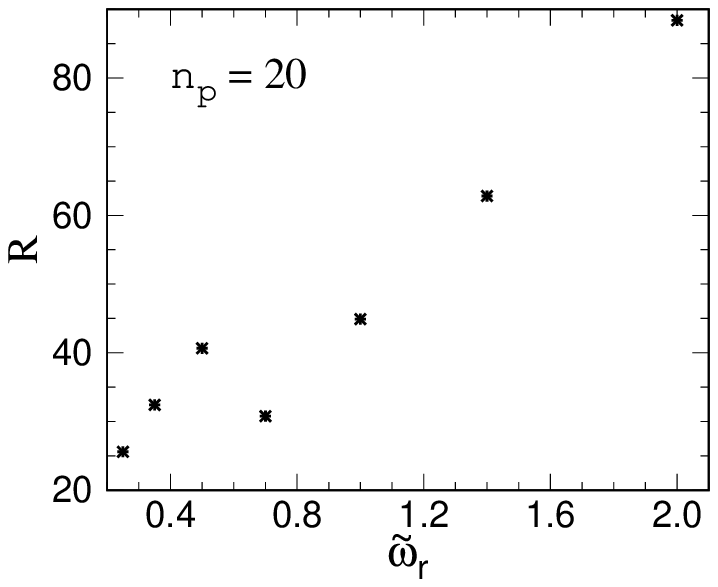}
\includegraphics{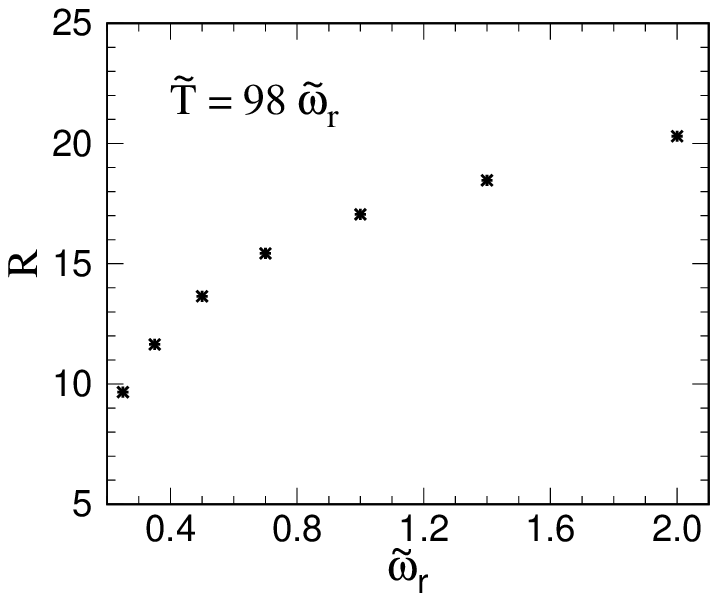}}
\caption{\label{FigMer}
The relative SWAP efficiency,
$R\equiv (\langle KE\rangle_0-\langle KE\rangle_f)/(\langle KE\rangle_0 P_\ell)$ for ${\cal B}=0.02$. One calculation was for a single
initial momentum, $20\hbar k$, and the other was for a thermal distribution.
}
\end{figure}

One possible measure of the efficiency of SWAP cooling is the change in
energy divided by the population lost to the leak state. In order
to make it dimensionless, we define it as
$R\equiv (\langle KE\rangle_0-\langle KE\rangle_f)/(\langle KE\rangle_0 P_\ell)$.
Larger $R$ implies a more effective SWAP either by having
a larger change in energy or smaller loss.
This quantity
is inversely proportional to ${\cal B}$ for small branching ratio
to the leak state which leads to the obvious conclusion that 
smaller loss is better.

Less obvious is the trend with respect to
the linewidth keeping the branching ratio small and fixed.
Figure~\ref{FigMer} shows the results of two calculations of the
relative efficiency with ${\cal B}=0.02$ as a function of the
scaled recoil frequency, $\tilde{\omega}_r$ (smaller decay rate,
$\gamma$, means larger $\tilde{\omega}_r$).
In the left graph, the calculation was performed
using the parameters of Fig.~\ref{FigLeakV} (i.e. a specific initial
momentum to mimic a beam) while the right graph used parameters of
Fig.~\ref{FigLeakT} (i.e. a thermal distribution). To give an idea
of the size that might be expected, the beam case would have
$R=0.19/0.02 = 9.5$ for Doppler cooling.

For a given
scaled recoil energy, SWAP was more efficient for the beam case than
for the thermal case, which is mainly due to the relative slope in
Fig.~\ref{FigLeakV} versus~\ref{FigLeakT}.
For the thermal case, the efficiency monotonically increases
with increasing $\tilde{\omega}_r$ which supports the proposition
that SWAP is better for narrow transitions. However, the increase is
not quite as rapid for larger $\tilde{\omega}_r$ which indicates
there may be limits to the effectiveness of cooling a thermal gas.
For the beam, there is a dip in efficiency at $\tilde{\omega}_r=0.7$
which matches the increase loss seen in Fig.~\ref{FigLeakV}. Otherwise,
the trend is an increase in efficiency with increasing
$\tilde{\omega}_r$ as expected; at $\tilde{\omega}_r=2$,
the relative efficiency is almost $10\times$ that for Doppler cooling.

\subsection{Implication for $\bar{\rm H}$}

One of the motivations for the above studies was to explore whether
$\bar{\rm H}$ can be further cooled from the apparent limit of
Doppler cooling in the ALPHA trap: the average final energy in
the simulation was
$\bar{E}_f/k_B\sim 30$~mK\cite{DFR13}. This simulation was based on one
laser Doppler cooling on the 1S-2P transition at 121.6~nm.
The idea for further cooling is to excite the $\bar{\rm H}$ to the
metastable 2S state and attempt to laser cool on the 2S-3P or 2S-4P
transition. The best case is the 2S-4P transition which
has the larger photon momentum and smaller decay rate, meaning it will
have the larger $\tilde{\omega}_r$; the value,
$\tilde{\omega}_r\simeq 0.065$, is smaller than any of the values
simulated above. For this 
transition, the Doppler temperature $T_D\simeq 0.31$~mK.
Thus, the starting $k_BT/(\hbar\omega_r)$ is $5\times$ that in
Fig.~\ref{FigLeakT} assuming $T\sim 20$~mK\cite{DFR13}.
Unfortunately, the branching ratio to leak states is
${\cal B}\simeq 0.88$ for either transition. We simulated multiple
SWAPs until all of the population had leaked out of the cycling transition
and found a decrease of 16\% in the kinetic
energy. This is a large decrease considering how unfavorable the
conditions (i.e. large branching ratio and large 
linewidth). However, since populating the 2S state is very
difficult, the simulations strongly suggest that SWAP cooling is not
worthwhile for further cooling $\bar{\rm H}$.



\section{Conclusion}

Results from simulations of
SWAP cooling of a thermal distribution or slowing a beam
were presented when a loss channel is present. For the cases investigated,
there was not a large difference in the cooling or slowing versus the
branching ratio, ${\cal B}$, for spontaneous decay into the leak
state(s) for one SWAP, and there was less population lost during one
SWAP when slowing a beam compared to cooling a thermal
distribution. The population lost to the leak state
strongly depended on the branching ratio as well as the spontaneous
decay rate of the excited state. For small branching ratios, the population
lost to the leak state is proportional to the branching ratio but
does not increase as rapidly for larger ${\cal B}$. For the same branching
ratio, atoms with larger scaled recoil energy, $\varepsilon_r/(\hbar\gamma )$,
have less population lost into the leak during each SWAP. For the cases
we investigated, the population lost during each SWAP for
small ${\cal B}$ was as much as
$5\times$ smaller than would be lost during Doppler cooling. This confirms
the suggestion in Ref.~\cite{NCB18} that SWAP could be useful for
cooling molecules. SWAP might be useful for slowing beams even for
unfavorable (i.e. large) branching ratios out of the cycling transition
if the scaled recoil energy is large.

This work was partially supported by NSF REU grant PHY-1852501 (JG and MEP)
and NSF grant PHY-1806380 (JG, MEP, and FR).

\bibliography{swap_w_leak}

\begin{thebibliography}{11}%
\makeatletter
\providecommand \@ifxundefined [1]{%
 \@ifx{#1\undefined}
}%
\providecommand \@ifnum [1]{%
 \ifnum #1\expandafter \@firstoftwo
 \else \expandafter \@secondoftwo
 \fi
}%
\providecommand \@ifx [1]{%
 \ifx #1\expandafter \@firstoftwo
 \else \expandafter \@secondoftwo
 \fi
}%
\providecommand \natexlab [1]{#1}%
\providecommand \enquote  [1]{``#1''}%
\providecommand \bibnamefont  [1]{#1}%
\providecommand \bibfnamefont [1]{#1}%
\providecommand \citenamefont [1]{#1}%
\providecommand \href@noop [0]{\@secondoftwo}%
\providecommand \href [0]{\begingroup \@sanitize@url \@href}%
\providecommand \@href[1]{\@@startlink{#1}\@@href}%
\providecommand \@@href[1]{\endgroup#1\@@endlink}%
\providecommand \@sanitize@url [0]{\catcode `\\12\catcode `\$12\catcode
  `\&12\catcode `\#12\catcode `\^12\catcode `\_12\catcode `\%12\relax}%
\providecommand \@@startlink[1]{}%
\providecommand \@@endlink[0]{}%
\providecommand \url  [0]{\begingroup\@sanitize@url \@url }%
\providecommand \@url [1]{\endgroup\@href {#1}{\urlprefix }}%
\providecommand \urlprefix  [0]{URL }%
\providecommand \Eprint [0]{\href }%
\providecommand \doibase [0]{http://dx.doi.org/}%
\providecommand \selectlanguage [0]{\@gobble}%
\providecommand \bibinfo  [0]{\@secondoftwo}%
\providecommand \bibfield  [0]{\@secondoftwo}%
\providecommand \translation [1]{[#1]}%
\providecommand \BibitemOpen [0]{}%
\providecommand \bibitemStop [0]{}%
\providecommand \bibitemNoStop [0]{.\EOS\space}%
\providecommand \EOS [0]{\spacefactor3000\relax}%
\providecommand \BibitemShut  [1]{\csname bibitem#1\endcsname}%
\let\auto@bib@innerbib\@empty
\bibitem [{\citenamefont {Metcalf}\ and\ \citenamefont {van~der
  Straten}(1999)}]{HJM99}%
  \BibitemOpen
  \bibfield  {author} {\bibinfo {author} {\bibfnamefont {H.~J.}\ \bibnamefont
  {Metcalf}}\ and\ \bibinfo {author} {\bibfnamefont {P.}~\bibnamefont {van~der
  Straten}},\ }\href@noop {} {\emph {\bibinfo {title} {Laser Cooling and
  Trapping}}}\ (\bibinfo  {publisher} {Springer},\ \bibinfo {year}
  {1999})\BibitemShut {NoStop}%
\bibitem [{\citenamefont {Foot}(2005)}]{CJF05}%
  \BibitemOpen
  \bibfield  {author} {\bibinfo {author} {\bibfnamefont {C.~J.}\ \bibnamefont
  {Foot}},\ }\href@noop {} {\emph {\bibinfo {title} {Classical Electrodynamics,
  3rd Edition}}}\ (\bibinfo  {publisher} {Oxford University Press},\ \bibinfo
  {year} {2005})\BibitemShut {NoStop}%
\bibitem [{\citenamefont {Carr}\ \emph {et~al.}(2009)\citenamefont {Carr},
  \citenamefont {DeMille}, \citenamefont {Krems},\ and\ \citenamefont
  {Ye}}]{CDK09}%
  \BibitemOpen
  \bibfield  {author} {\bibinfo {author} {\bibfnamefont {L.~D.}\ \bibnamefont
  {Carr}}, \bibinfo {author} {\bibfnamefont {D.}~\bibnamefont {DeMille}},
  \bibinfo {author} {\bibfnamefont {R.~V.}\ \bibnamefont {Krems}}, \ and\
  \bibinfo {author} {\bibfnamefont {J.}~\bibnamefont {Ye}},\ }\bibfield
  {title} {\enquote {\bibinfo {title} {Cold and ultracold molecules: science,
  technology and applications},}\ }\href@noop {} {\bibfield  {journal}
  {\bibinfo  {journal} {New J. Phys.}\ }\textbf {\bibinfo {volume} {11}},\
  \bibinfo {pages} {055049} (\bibinfo {year} {2009})}\BibitemShut {NoStop}%
\bibitem [{\citenamefont {Novotny}\ and\ \citenamefont {Hecht}(2012)}]{LNH12}%
  \BibitemOpen
  \bibfield  {author} {\bibinfo {author} {\bibfnamefont {L.}~\bibnamefont
  {Novotny}}\ and\ \bibinfo {author} {\bibfnamefont {B.}~\bibnamefont
  {Hecht}},\ }\href@noop {} {\emph {\bibinfo {title} {Principles of
  nano-optics}}}\ (\bibinfo  {publisher} {Cambridge University Press},\
  \bibinfo {year} {2012})\BibitemShut {NoStop}%
\bibitem [{\citenamefont {Norcia}\ \emph {et~al.}(2018)\citenamefont {Norcia},
  \citenamefont {Cline}, \citenamefont {Bartolotta}, \citenamefont {Holland},\
  and\ \citenamefont {Thompson}}]{NCB18}%
  \BibitemOpen
  \bibfield  {author} {\bibinfo {author} {\bibfnamefont {M.~A.}\ \bibnamefont
  {Norcia}}, \bibinfo {author} {\bibfnamefont {J.~R.~K.}\ \bibnamefont
  {Cline}}, \bibinfo {author} {\bibfnamefont {J.~P.}\ \bibnamefont
  {Bartolotta}}, \bibinfo {author} {\bibfnamefont {M.~J.}\ \bibnamefont
  {Holland}}, \ and\ \bibinfo {author} {\bibfnamefont {J.~K.}\ \bibnamefont
  {Thompson}},\ }\bibfield  {title} {\enquote {\bibinfo {title} {Narrow-line
  laser cooling by adiabatic transfer},}\ }\href@noop {} {\bibfield  {journal}
  {\bibinfo  {journal} {New J. Phys.}\ }\textbf {\bibinfo {volume} {20}},\
  \bibinfo {pages} {023021} (\bibinfo {year} {2018})}\BibitemShut {NoStop}%
\bibitem [{\citenamefont {Bartolotta}\ \emph {et~al.}(2018)\citenamefont
  {Bartolotta}, \citenamefont {Norcia}, \citenamefont {Cline}, \citenamefont
  {Thompson},\ and\ \citenamefont {Holland}}]{BNC18}%
  \BibitemOpen
  \bibfield  {author} {\bibinfo {author} {\bibfnamefont {J.~P.}\ \bibnamefont
  {Bartolotta}}, \bibinfo {author} {\bibfnamefont {M.~A.}\ \bibnamefont
  {Norcia}}, \bibinfo {author} {\bibfnamefont {J.~R.~K.}\ \bibnamefont
  {Cline}}, \bibinfo {author} {\bibfnamefont {J.~K.}\ \bibnamefont {Thompson}},
  \ and\ \bibinfo {author} {\bibfnamefont {M.~J.}\ \bibnamefont {Holland}},\
  }\bibfield  {title} {\enquote {\bibinfo {title} {Laser cooling by
  sawtooth-wave adiabatic passage},}\ }\href@noop {} {\bibfield  {journal}
  {\bibinfo  {journal} {Phys. Rev. A}\ }\textbf {\bibinfo {volume} {98}},\
  \bibinfo {pages} {023404} (\bibinfo {year} {2018})}\BibitemShut {NoStop}%
\bibitem [{\citenamefont {Petersen}\ \emph {et~al.}(2019)\citenamefont
  {Petersen}, \citenamefont {M{\"u}hlbauer}, \citenamefont {Bougas},
  \citenamefont {Sharma}, \citenamefont {Budker},\ and\ \citenamefont
  {Windpassinger}}]{PMB19}%
  \BibitemOpen
  \bibfield  {author} {\bibinfo {author} {\bibfnamefont {N.}~\bibnamefont
  {Petersen}}, \bibinfo {author} {\bibfnamefont {F.}~\bibnamefont
  {M{\"u}hlbauer}}, \bibinfo {author} {\bibfnamefont {L.}~\bibnamefont
  {Bougas}}, \bibinfo {author} {\bibfnamefont {A.}~\bibnamefont {Sharma}},
  \bibinfo {author} {\bibfnamefont {D.}~\bibnamefont {Budker}}, \ and\ \bibinfo
  {author} {\bibfnamefont {P.}~\bibnamefont {Windpassinger}},\ }\bibfield
  {title} {\enquote {\bibinfo {title} {Sawtooth-wave adiabatic-passage slowing
  of dysprosium},}\ }\href@noop {} {\bibfield  {journal} {\bibinfo  {journal}
  {Phys. Rev. A}\ }\textbf {\bibinfo {volume} {99}},\ \bibinfo {pages} {063414}
  (\bibinfo {year} {2019})}\BibitemShut {NoStop}%
\bibitem [{\citenamefont {Snigirev}\ \emph {et~al.}(2019)\citenamefont
  {Snigirev}, \citenamefont {Park}, \citenamefont {Heinz}, \citenamefont
  {Bloch},\ and\ \citenamefont {Blatt}}]{SPH19}%
  \BibitemOpen
  \bibfield  {author} {\bibinfo {author} {\bibfnamefont {S.}~\bibnamefont
  {Snigirev}}, \bibinfo {author} {\bibfnamefont {A.~J.}\ \bibnamefont {Park}},
  \bibinfo {author} {\bibfnamefont {A.}~\bibnamefont {Heinz}}, \bibinfo
  {author} {\bibfnamefont {I.}~\bibnamefont {Bloch}}, \ and\ \bibinfo {author}
  {\bibfnamefont {S.}~\bibnamefont {Blatt}},\ }\bibfield  {title} {\enquote
  {\bibinfo {title} {Fast and dense magneto-optical traps for strontium},}\
  }\href@noop {} {\bibfield  {journal} {\bibinfo  {journal} {Phys. Rev. A}\
  }\textbf {\bibinfo {volume} {99}},\ \bibinfo {pages} {063421} (\bibinfo
  {year} {2019})}\BibitemShut {NoStop}%
\bibitem [{\citenamefont {Liang}\ \emph {et~al.}(2019)\citenamefont {Liang},
  \citenamefont {Chen}, \citenamefont {Bu}, \citenamefont {Zhang},\ and\
  \citenamefont {Yan}}]{LCB19}%
  \BibitemOpen
  \bibfield  {author} {\bibinfo {author} {\bibfnamefont {Q.}~\bibnamefont
  {Liang}}, \bibinfo {author} {\bibfnamefont {T.}~\bibnamefont {Chen}},
  \bibinfo {author} {\bibfnamefont {W.}~\bibnamefont {Bu}}, \bibinfo {author}
  {\bibfnamefont {Y.}~\bibnamefont {Zhang}}, \ and\ \bibinfo {author}
  {\bibfnamefont {B.}~\bibnamefont {Yan}},\ }\bibfield  {title} {\enquote
  {\bibinfo {title} {Laser cooling with adiabatic passage for diatomic
  molecules},}\ }\href@noop {} {\bibfield  {journal} {\bibinfo  {journal}
  {arXiv preprint arXiv:1902.05212}\ } (\bibinfo {year} {2019})}\BibitemShut
  {NoStop}%
\bibitem [{\citenamefont {Donnan}\ \emph {et~al.}(2013)\citenamefont {Donnan},
  \citenamefont {Fujiwara},\ and\ \citenamefont {Robicheaux}}]{DFR13}%
  \BibitemOpen
  \bibfield  {author} {\bibinfo {author} {\bibfnamefont {P.~H.}\ \bibnamefont
  {Donnan}}, \bibinfo {author} {\bibfnamefont {M.~C.}\ \bibnamefont
  {Fujiwara}}, \ and\ \bibinfo {author} {\bibfnamefont {F.}~\bibnamefont
  {Robicheaux}},\ }\bibfield  {title} {\enquote {\bibinfo {title} {A proposal
  for laser cooling antihydrogen atoms},}\ }\href@noop {} {\bibfield  {journal}
  {\bibinfo  {journal} {J. Phys. B}\ }\textbf {\bibinfo {volume} {46}},\
  \bibinfo {pages} {025302} (\bibinfo {year} {2013})}\BibitemShut {NoStop}%
\bibitem [{\citenamefont {Ahmadi}\ and\ \citenamefont {et~al
  (ALPHA~collaboration)}(2018)}]{ALP18}%
  \BibitemOpen
  \bibfield  {author} {\bibinfo {author} {\bibfnamefont {M}~\bibnamefont
  {Ahmadi}}\ and\ \bibinfo {author} {\bibnamefont {et~al
  (ALPHA~collaboration)}},\ }\bibfield  {title} {\enquote {\bibinfo {title}
  {Characterization of the 1s--2s transition in antihydrogen},}\ }\href@noop {}
  {\bibfield  {journal} {\bibinfo  {journal} {Nature}\ }\textbf {\bibinfo
  {volume} {557}},\ \bibinfo {pages} {71} (\bibinfo {year} {2018})}\BibitemShut
  {NoStop}%
\end{thebibliography}%

\end{document}